\documentclass[aps,pra,showkeys,twocolumn,superscriptaddress]{revtex4-2}
\usepackage{array}
\usepackage{booktabs}
\usepackage{tabu}
\usepackage{dcolumn}
\usepackage{floatrow}
\usepackage{amsmath}
\usepackage{amsfonts}
\usepackage{float}
\usepackage{amssymb}
\usepackage{graphicx,color}
\usepackage[colorlinks={true}]{hyperref}
\hypersetup{colorlinks=true,linkcolor=red,citecolor=blue,urlcolor=blue}
\usepackage{graphicx}
\usepackage{dcolumn}
\usepackage{bm}
\usepackage{pstricks}
\usepackage{braket}
\usepackage{orcidlink}

\def\be{\begin{equation}}
  \def\ee{\end{equation}}
\def\bea{\begin{eqnarray}}
\def\eea{\end{eqnarray}}
\def\f{\frac}
\def\n{\nonumber}
\def\l{\label}
\def\p{\phi}
\def\o{\over}
\def\R{\rho}
\def\pa{\partial}
\def\om{\omega}
\def\na{\nabla}
\def\P{\Phi}
\begin{document}

\title{Work extraction from quantum coherence in non-equilibrium environment} 

\author{Maryam Hadipour \orcidlink{0000-0002-6573-9960}}
\affiliation{Faculty of Physics, Urmia University of Technology, Urmia, Iran}

\author{Soroush Haseli \orcidlink{0000-0003-1031-4815}}\email{soroush.haseli@uut.ac.ir}
\affiliation{Faculty of Physics, Urmia University of Technology, Urmia, Iran}

\date{\today}
\def\be{\begin{equation}}
  \def\ee{\end{equation}}
\def\bea{\begin{eqnarray}}
\def\eea{\end{eqnarray}}
\def\f{\frac}
\def\n{\nonumber}
\def\l{\label}
\def\p{\phi}
\def\o{\over}
\def\R{\rho}
\def\pa{\partial}
\def\om{\omega}
\def\na{\nabla}
\def\P{$\Phi$}

\begin{abstract}
Ergotropy, as a measure for extractable work from a quantum system, has garnered significant attention due to its relevance in quantum thermodynamics and information processing. In this work, the dynamics of ergotropy will be investigated in a nonequilibrium environment for both Markovian and non-Markovian regime. In this study, both the coherent and incoherent parts of  the ergotropy will be considered. It will be shown that for a non-equilibrium environment, the extraction of work is more efficient compared to when the environment is in equilibrium.  
\end{abstract}

\maketitle

\section{Introduction}
Quantum thermodynamics investigates the interplay between the macroscopic principles of thermodynamics and the microscopic world governed by quantum mechanics. It plays a crucial role in understanding the fundamental energetic constraints of quantum technologies. Since many quantum tasks involve energy transfer processes, quantum thermodynamics provides a framework to quantify the maximum extractable work, termed ergotropy \cite{1}, from a quantum system via unitary operations. Restricted to unitary operations, ergotropy embodies the process of extracting energy in the form of work, facilitated by the presence of an external field.  On the other hand, the interaction of the system with the environment will lead to a change in the level of ergotropy. In the framework of open quantum systems \cite{2}, heat and work can interchange via thermodynamic processes. Such interchange facilitates the development and experimentation of diverse quantum devices, encompassing quantum heat engines \cite{3,4,5}, quantum refrigerators \cite{6}, quantum batteries \cite{7,s1,s2,s3,s4,s5,s6,8,9,10,11,12,13,14,15,16,17,18,19,20,21,22,23,24,25,26,27,28,29,30,31,32,33,34,35,36,37,38,39,40,41,42,43,44,45,46,47,48}.
Researchers are investigating the potential of ergotropy as a resource for executing various quantum tasks, including quantum computing, quantum cryptography, and quantum simulation. The use of ergotropy as a resource for quantum applications must adhere to the energy balance dictated by the first law of thermodynamics.  Alicki has introduced the quantum version of the first law of thermodynamics \cite{49}. In this description, the internal energy is considered to be the average value of the Hamiltonian of the quantum system, which governs the transformations.  The work,  is associated with a change in the gap structure of the energy spectrum induced by a time-dependent Hamiltonian. 

Identifying uniquely quantum signatures in thermodynamic settings has been a central theme in the field of quantum thermodynamics over the past decade. This includes the identification of quantum signatures in various aspect such as thermal machines \cite{50,51,52,53,54,55,56,57,58,59,60,61,62,63,64,65}, work extraction protocols \cite{65,66,67,68,69,70,71,72,73,74,75,76,77,78,79}, in fluctuations of work \cite{80,81,82,83,84,85,86}, and in work
deposition processes \cite{87,88,89,90,91,92,93}. In the context of quantum thermodynamics, a quantum signature refers to distinctive features or phenomena that arise due to the inherently quantum nature of the system under consideration. One could confidently assert that quantum coherence is the most fundamental of all non-classical features. However, precise mathematical techniques for its quantification have only recently been developed within the field of quantum information theory \cite{94,95}. So far, many studies have focused on highlighting the role of quantum coherence from the perspective of quantum thermodynamics \cite{96,97,98,99,100,101,102,103,104,105,106}. In quantum information theory, coherence is a basis-dependent quantity that can be geometrically expressed as the distance between the state of the system and its dephased counterpart \cite{107}. The notion of the quantum coherence establishes a connection with the finite-time thermodynamics of quantum systems, where relative entropy is widely used to evaluate the irreversible entropy production   in closed and open quantum systems \cite{108,109,110,111,112,113,114,115,116,117,118}. Recently, this connection has been utilized to distinguish the coherent component of entropy production in quantum dynamics \cite{119,120,121,122}. In recent decades, there has been a growing theoretical interest in studying the dynamics of open quantum systems\cite{2}. The dynamics of ergotropy in the presence of an external environment has been a subject of much research in recent years \cite{41,123,124,125}. It is worth noting that in all studies conducted so far, the environment has been assumed to be in equilibrium with stable statistical properties, and the non-equilibrium effects of the environment on ergotropy have not been investigated yet. Recently, the nonequilibrium feature of environments has been experimentally observed in many vital dynamical processes. In such processes, the initial state of the environment, which is disturbed from equilibrium due to the interaction with the system, cannot reach equilibrium over time. Consequently, the environment surrounding the quantum system remains out of equilibrium \cite{126,127,128,129,130,131,132}. Furthermore, it has been shown in recent years that the non-equilibrium feature of the environment with unstable statistical properties leads to a frequency shift, which will have a significant effect on reducing decoherence effects \cite{133,134,135,136}.
In this work, by considering both the coherent and incoherent contributions of the ergotropy, we investigate the dynamics of ergotropy in non-equilibrium environments. In the studies conducted in this work, both Markovian and non-Markovian regimes will be considered. The work is organized as follow.
\section{THEORETICAL FRAMEWORK}
\subsection{THE FULL ERGOTROPY, ITS COHERENT AND INCOHERENT PARTS}
Ergotropy refers to the maximum energy extractable from a quantum system through cyclic unitary process \cite{1}. This cyclic unitary process can be characterized by the  unitary transformation operation $U$.  
The full  ergotropy can be obtained as $\mathcal{E}(\rho)=Tr(\rho H)-\min_{U}Tr(H U \rho U^{\dag})$, where $H$ is the Hamiltonian of the system and the minimum is taken over the set of all unitary transformations. It has been shown that for any given arbitrary  state, there exists a unique state that maximizes the above equation. This state is referred to as the passive state $\hat{P}_\rho$. Thus the full ergotropy can be obtained as
\begin{equation}
\mathcal{E}(\rho)=Tr \left\lbrace  H(\hat{\rho}-\hat{P}_\rho)\right\rbrace. 
\end{equation}

Considering the spectral decomposition of the density matrix $\rho$ and the Hamiltonian $H$ as 
\begin{eqnarray}
\rho&=&\sum_n r_n \vert r_n \rangle \langle r_n \vert, \quad r_1 \geq r_2 \geq ... \geq r_n \\  
H&=&\sum_m \varepsilon_{m} \vert \varepsilon_m \rangle \langle \varepsilon_m \vert  \quad \varepsilon_1 \leq \varepsilon_2 \leq ... \leq \varepsilon_m,
\end{eqnarray}
the passive state can be obtained as $\hat{P}_\rho= \sum_n r_n \vert \varepsilon_n \rangle \langle \varepsilon_n \vert$, where $r_n$ ($\vert r_n \rangle$) and  $\varepsilon_m$ ($\vert \varepsilon_m \rangle$) are eigenvalues (eigenstates) of the density matrix $\rho$ and $H$ respectively. So, the ergotropy $\mathcal{E}(\rho)$ is obtained as
\begin{equation}\label{tergo}
\mathcal{E}(\rho)= \sum_{n,m} r_n \varepsilon_m \left( \left| \langle r_n \vert \varepsilon_m \rangle \right|^2 - \delta_{m,n}  \right), 
\end{equation}
where $\delta_{m,n}$ is the Kronecker delta function.  Now, let's examine the concept of the incoherent part of ergotropy. Let's begin by exploring the incoherent component of ergotropy $\mathcal{E}_i(\rho)$. $\mathcal{E}_i(\rho)$ can be considered as the maximum work that can be extracted from $\hat{\rho}$ without altering its coherence. Actually, $\mathcal{E}_i(\rho)$  denotes the maximum work that can be extracted from $\hat{\rho}$ after destroying  all coherence of the state using a phase-damping map $\Delta$.  $\mathcal{E}_i(\rho)$ can be regarded as the ergotropy of the incoherent state $\hat{\delta}_\rho$, which has the same energy distribution as $\hat{\rho}$ with zero coherence.  To calculate the incoherent ergotropy, we need to first determine the passive state corresponding to the dephased state $\hat{\delta}_\rho$, which is denoted as $\hat{P}_\delta$. So, the incoherent part of the ergotropy can be obtained as
\begin{equation}\label{inc}
\mathcal{E}_i(\rho) \equiv \mathcal{E}(\hat{\delta}_\rho)= Tr \left\lbrace \hat{H}(\hat{\delta}_\rho-\hat{P}_\delta)\right\rbrace .
\end{equation}
Once the incoherent part of ergotropy is determined, the coherent part can be easily obtained as 
\begin{equation}\label{coh}
\mathcal{E}_c(\rho)=\mathcal{E}(\rho)-\mathcal{E}_i(\rho).
\end{equation}
\subsection{Dynamics of ergotropy under nonequilibrium
decoherence processes}
Let us consider a two-level quantum system interacting with a non-equilibrium environment. The effect of the environment on the quantum system can be explained through  stochastic fluctuations in a system observable, which is described by the Kubo-Anderson spectral diffusion process \cite{137,138,139}. In this context, it is assumed that the system's energy remains constant, with decoherence of the quantum system being solely induced by the stochastic fluctuations. The pure decoherence Hamiltonian for the system is formulated as \cite{140,141}
\begin{equation}
H(t)=H_0 + \mathcal{V}(t)=\frac{\hbar}{2} \omega_0 \sigma_z +\frac{\hbar}{2} \zeta(t)\sigma_z  ,
\end{equation}
where $\sigma_z$ is the z-component of Pauli operator, $\omega_0$ is the transition frequency between excited $\vert e \rangle$ and ground state $\vert g \rangle$.  $\zeta(t)$ describes the environmental noise with both non-stationary and non-Markovian properties. The time evolution of the total density matrix $\rho(t;\zeta(t))$ follows the Liouville equation
\begin{equation}
\frac{\partial}{\partial t}\rho(t;\zeta(t))=-\frac{i}{\hbar} \left[ H(t), \rho(t;\zeta(t)) \right].
\end{equation}
The total density matrix $\rho(t;\zeta(t))$ depends on the environmental noise induced by the non-equilibrium environment. The elements of the total density matrix in the basis set $\lbrace \vert e \rangle, \vert g \rangle \rbrace$ satisfy the following stochastic differential equations as
\begin{eqnarray}\label{df}
\dot{\rho}_{ee}(t;\zeta(t))&=&0 \nonumber \\
\dot{\rho}_{ge}(t;\zeta(t))&=& i\left[\omega_0 + \zeta(t) \right] \rho_{ge}(t;\zeta(t)).
\end{eqnarray}
For the other elements of the density matrix, the stochastic differential equations will be easily obtained through utilizing the properties of the density matrix, namely $Tr[\rho(t;\zeta(t))]=1$ and $\rho^{\dag}(t;\zeta(t))=\rho(t;\zeta(t))$. By integrating the differential equations in Eq.\ref{df}, we will have
\begin{eqnarray}
\rho_{ee}(t;\zeta(t))&=&\rho_{ee}(0;\zeta(0)),  \\
\rho_{ge}(t;\zeta(t))&=&\exp[i \omega_0 t+ i \int_0^t \zeta(t^{\prime})dt^\prime]\rho_{ge}(0;\zeta(0)). \nonumber
\end{eqnarray}
The reduced density matrix of the system is obtained  through taking an average over the environmental noise as $\rho(t)= \langle \rho(t;\zeta(t)) \rangle $. Assuming no initial correlation between the system and its environment, so the elements of the reduced density matrix can be obtained as 
\begin{eqnarray}
\rho_{ee}(t)&=&\rho_{ee}(0), \\
\rho_{ge}(t)&=&\exp[i \omega_0 t] F(t) \rho_{ge}(0),
\end{eqnarray}
where $\rho_{gg}(t)=1-\rho_{ee}(t)$, $\rho_{eg}(t)=\rho_{ge}^{\star}9t)$ and $F(t)=\langle \exp[i \int_0^t dt^\prime \zeta(t^\prime)]\rangle$ is the decoherence factor of the system. Here, the initial state of the system is considered  in the Bloch representation as $\rho(0)=\frac{1}{2}(\mathrm{I}+\sum_{i} r_i \sigma_i)$, where $r_i$ and $\sigma_i$, with $i \in \left\lbrace x,y,z \right\rbrace $ are the $i$th component of the  Bloch vector and Pauli operators respectively. So, the reduce density matrix of the quantum system at time $t$ can be obtained as 
\begin{equation}\label{Mat}
\rho(t)=\left(
\begin{array}{cc}
 \frac{1+r_z}{2}  & \frac{r_x-ir_y}{2} e^{- i\omega_0 t}F^{\star}(t)  \\
 \frac{r_x+ir_y}{2} e^{ i\omega_0 t}F(t) & \frac{1-r_z }{2} \\
\end{array}
\right).
\end{equation}
In the considered non-equilibrium process, the environmental noise $\zeta(t)$ includes a non-stationary and non-Markovian dichotomous process. The considered non-equilibrium decoherence process is described through non-equilibrium  parameter  $a$ and the memory decay rate $\kappa$. Additionally, its range randomly varies with the jumping rate $\lambda$ between two values $\pm v$ \cite{133,134,135,142}. In the context of non-equilibrium dynamical decoherence process, it is possible to analytically determine the decoherence factor $F(t)$ as \cite{133}
\begin{eqnarray}\label{f}
F(t)&=& \mathfrak{L}^{-1}\left[ \mathfrak{F}(s) \right], \\
\mathfrak{F}(s)&=&\frac{s^2+(\kappa + i a v)s+\kappa(2 \lambda+ ia v)}{s^3+\kappa s^2+(2 \kappa \lambda + v^2)s+\kappa v^2}. \nonumber
\end{eqnarray}
where $\mathfrak{L}$ is the inverse Laplace transform. The initial condition to obtain Eq.\ref{f} is $F(0)=1$. Before delving into the discussion regarding the dynamics of ergotropy under this non equilibrium stochastic process, it's preferable to address the nature of the process in terms of being Markovian or non-Markovian. In Markovian evolution, information flows continuously from the system to the environment, while in non-Markovian evolution, back-flow of information occurs from environment to system. In Markovian evolution, the future state of the system is  depend only on its current state, meaning the future is independent of past states, making the evolution memory-less. Conversely, in non-Markovian evolution, the system's future state depends on its previous states, thus making the evolution with-memory \cite{2}. So far, various criteria have been introduced for measuring the degree of non-Markovianity of a quantum process \cite{143,144,145,146,147,148,149,150}. Here, the coherence-based measure of non-Markovianity is used to quantify the degree of non-Markovianity of the non-equilibrium decoherence process \cite{150}. Quantum coherence is a measure of the superposition of states in a quantum system and can be quantified using the $l_1$ norm of coherence. For a quantum system that describes by density matrix $\rho$ the $l_1$ norm of coherence is defined by
\begin{equation}\label{coh}
C_{l_1}(\rho)=\sum_{i \neq j} \vert \rho_{ij} \vert. 
\end{equation}   
One method to measure non-Markovianity involves monitoring the dynamics of quantum coherence and detecting any departures from monotonic decay. Quantum coherence based measure of non-Markovianity can be written as 
\begin{equation}\label{non}
\mathcal{N}_c=\max_{\rho(0) \in \left\lbrace \vert \psi_d \rangle \right\rbrace } \int_{\frac{d C_{l_1}(\rho(t))}{dt}>0} \frac{d C_{l_1}(\rho(t))}{dt} dtt,
\end{equation}
where $\vert \psi_d \rangle = 1/\sqrt{d} \sum_{i=1}^{d} e^{i \theta_i} \vert i \rangle$, $d$ is the dimension of the Hilbert space and $\theta_i \in [0,2\pi)$. From Eqs.\ref{coh} and \ref{non}  the quantum coherence based measure of non-Markovianity for non-equilibrium decoherence process can be obtained as 
\begin{equation}
\mathcal{N}_c=\int_{\frac{d \vert F(t) \vert}{dt}>0}\frac{d \vert F(t) \vert}{dt}dt.
\end{equation}
\begin{figure}[H]
    \centering
  \includegraphics[width = 1\linewidth]{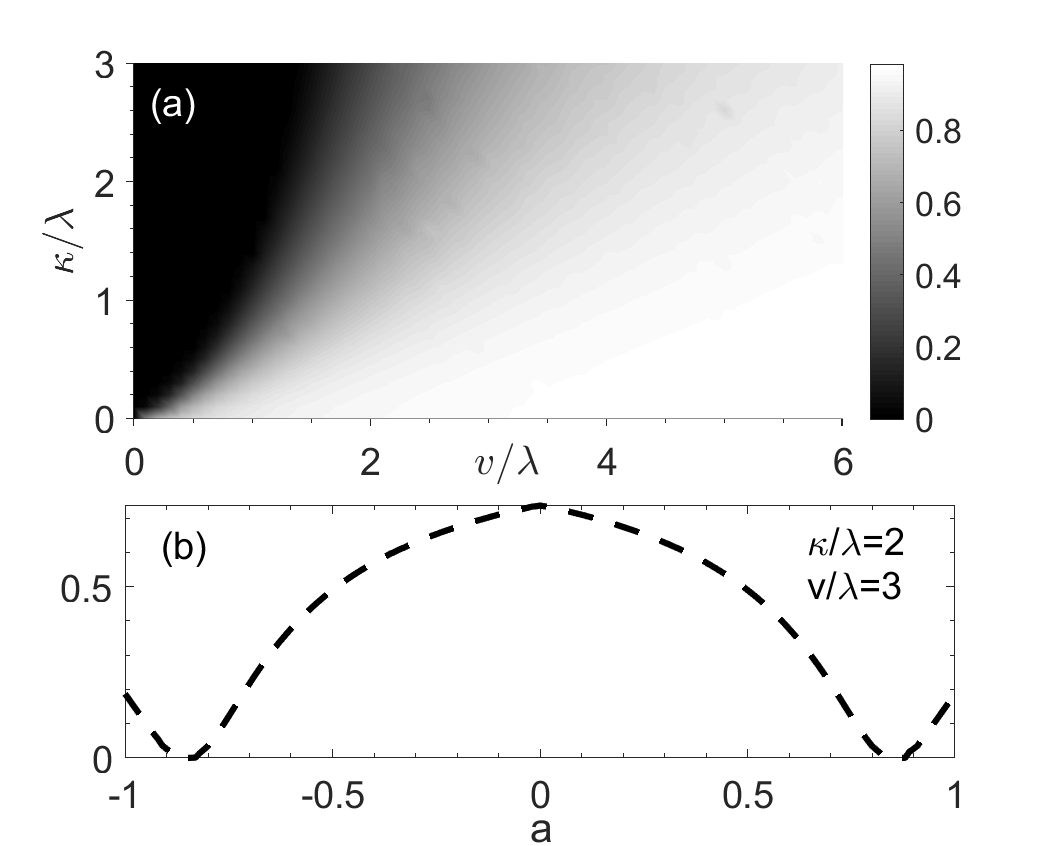}
    \centering
    \caption{(a)Non-Markovianity $\mathcal{N}_c$ for non-equilibrium decoherence process in terms of $\kappa / \lambda$ and $ v/\lambda $ with $a=0$, (b) Non-Markovianity $\mathcal{N}_c$ for non-equilibrium decoherence process in terms of $a$ with $\kappa / \lambda = 2 $ and $ v/\lambda =3 $.}\label{Fig1}
\end{figure}
Fig.\ref{Fig1}(a) shows the non-Markovianity $\mathcal{N}_c$  as functions of coupling $v/\lambda$ and the memory decay rate $\kappa/\lambda$ for $a=0$. As can be seen in strong coupling regime the dynamics is non-Markovian even for amall value of memory decay rate. In Fig.\ref{Fig1}(b), the effect of the non-equilibrium parameter $a$ on the degree of non-Markovianity is shown. To show the effect of $a$ on non-Markovianity from Fig.\ref{Fig1}(a) we select two values $\kappa/\lambda = 2$ and $v/\lambda =3$ such that even with $a=0$, the evolution remains non-Markovian. It can be seen that the degree of non-Markovianity decreses with increasing non-equilibrium parameter $a$. If the values of $\kappa / \lambda$ and $v / \lambda$ are chosen using Fig.\ref{Fig1}(a) such that the evolution is Markovian for $a=0$, then changing $a$ will not affect the degree of non-Markovianity and the dynamics is Markovian for all values of $a$. For instance, with $V/\lambda=0.8$ and $\kappa / \lambda=2$, the evolution is Markovian, and even with changes in $a$, the degree of non-Markovianity remains zero.  After revealing the non-Markovian range of non-equilibrium transformation, we return to the issue of ergotropy dynamics in both Markovian and non-Markovian regimes.

By examining the evolution from the perspective of memory effects (Markovian and non-Markovian), we now return to the topic of ergotropic dynamic. We will investigate the influence of a non-equilibrium environment on ergotropy for both Markovian and non-Markovian regimes. From Eqs.\ref{tergo} and \ref{Mat}, the total ergotropy $\mathcal{E}(\rho)$ can be obtained as
\begin{equation}
\mathcal{E}(\rho)=\frac{1}{2}\left(r_z + \sqrt{(r^{2}_x+r^{2}_y)\vert F(t) \vert^{2} + r_{z}^2} \right). 
\end{equation} 
The above relation can be rewritten as $\mathcal{E}(\rho)=\mathcal{U(\rho)}+\frac{1}{2}\sqrt{C_{l_1}(\rho)^{2}+4 \mathcal{U}(\rho)^{2}}$, by substituting the relations $C_{l_1}(\rho)=\sqrt{r_x^2+r_y^2}\vert F(t) \vert$  and $\mathcal{U}(\rho)=Tr[\rho(t)H_0]=r_z/2$, where $C_{l_1}(\rho(t))$ and $\mathcal{U}(\rho(t))$ are the $l_1$ norm of coherence and the internal energy of the system at time $t$ respectively. So, it can be concluded that the total ergotropy depends on quantum coherence and internal energy of the quantum system. As previously mentioned, the total ergotropy includes contributions from both the coherent and incoherent parts. From Eq.\ref{inc}, the incoherent part of ergotropy can be obtain as 
\begin{equation}
\mathcal{E}_i(\rho)=2 \max \left\lbrace 0, \mathcal{U}(\rho)\right\rbrace = \max \left\lbrace 0,r_z\right\rbrace. 
\end{equation}
Using Eq. \ref{coh}, the contribution of the coherent part of the ergotropy is obtained  as 
\begin{eqnarray}\label{cop}
\mathcal{E}_c(\rho)&=&\frac{1}{2}\left( \sqrt{(r^{2}_x+r^{2}_y)\vert F(t) \vert^{2} + r_{z}^2} -r_z \right) \nonumber \\
&=&\frac{1}{2}\sqrt{C_{l_1}(\rho)^{2}+4 \mathcal{U}(\rho)^{2}}-\mathcal{U(\rho)}.
\end{eqnarray}
\begin{figure}[H]
    \centering
  \includegraphics[width = 1\linewidth]{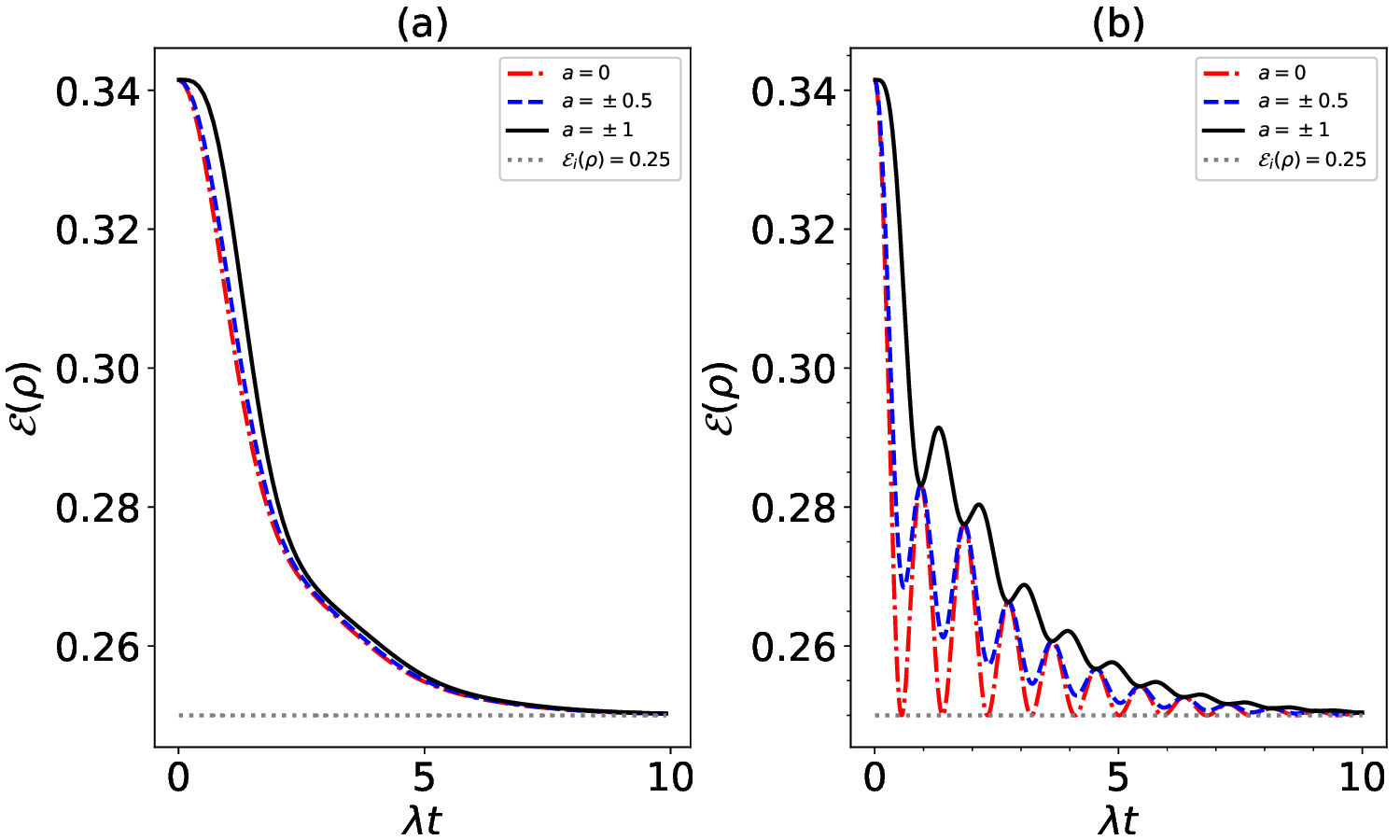}
    \centering
    \caption{(a)Total ergotropy $\mathcal{E}(\rho)$  and $\mathcal{E}_i(\rho)$ as a function of dimensionless parameter $\lambda t$ for different amount of the non-equilibrium parameter  $a$ with $r_x=r_y=r_z=0.25$ in (a) Markovian regime, $v/\lambda=0.8$ and $\kappa/\lambda =2$. (b)non-Markovian regime, $v/\lambda=3$ and $\kappa/\lambda =2$.}\label{Fig2}
\end{figure}
In Fig.\ref{Fig2}, the total ergotropy  has been plotted as a function of dimensionless parameter $\lambda t$ for both Markovian and non-Markovian regime for different values of the non-equilibrium  parameter $a$. The initial state of the system is considered  with the components of the Bloch vector as ($r_x=r_y=r_z=0.25$).

As can be seen for both Markovian Fig.\ref{Fig2}(a) and non-Markovian Fig.\ref{Fig2}(b) regime the total ergotropy $\mathcal{E}(\rho)$ increases with increasing the value of the non-equilibrium  parameter $a$. In other words it can be said that the ergotropy is enhanced as the environment away from equilibrium. However, the incoherent part of the ergotropy $\mathcal{E}_i(\rho)$ does not change due to the nature of the considered non-equilibrium decoherence process. The considered non-equilibrium decoherence process  similar to dephasing noise  primarily affects the coherence of the quantum system and not the population of energy levels. So, according to the fact that the incoherent part of the ergotropy depends only on the internal energy of the system, it can be said that the lack of change in the incoherent part is due to the nature of the considered noise. From Fig.\ref{Fig2}(a) and Fig.\ref{Fig2}(b), it can be observed that for both Markovian and non-Markovian regimes, as time progresses and coherence diminishes, the total ergotropy tends asymptotically towards the incoherent part of the ergotropy, i.e. $\lim_{t \to \infty} \mathcal{E}(\rho) = \mathcal{E}_i(\rho)$. 
\begin{figure}[H]
    \centering
  \includegraphics[width = 1\linewidth]{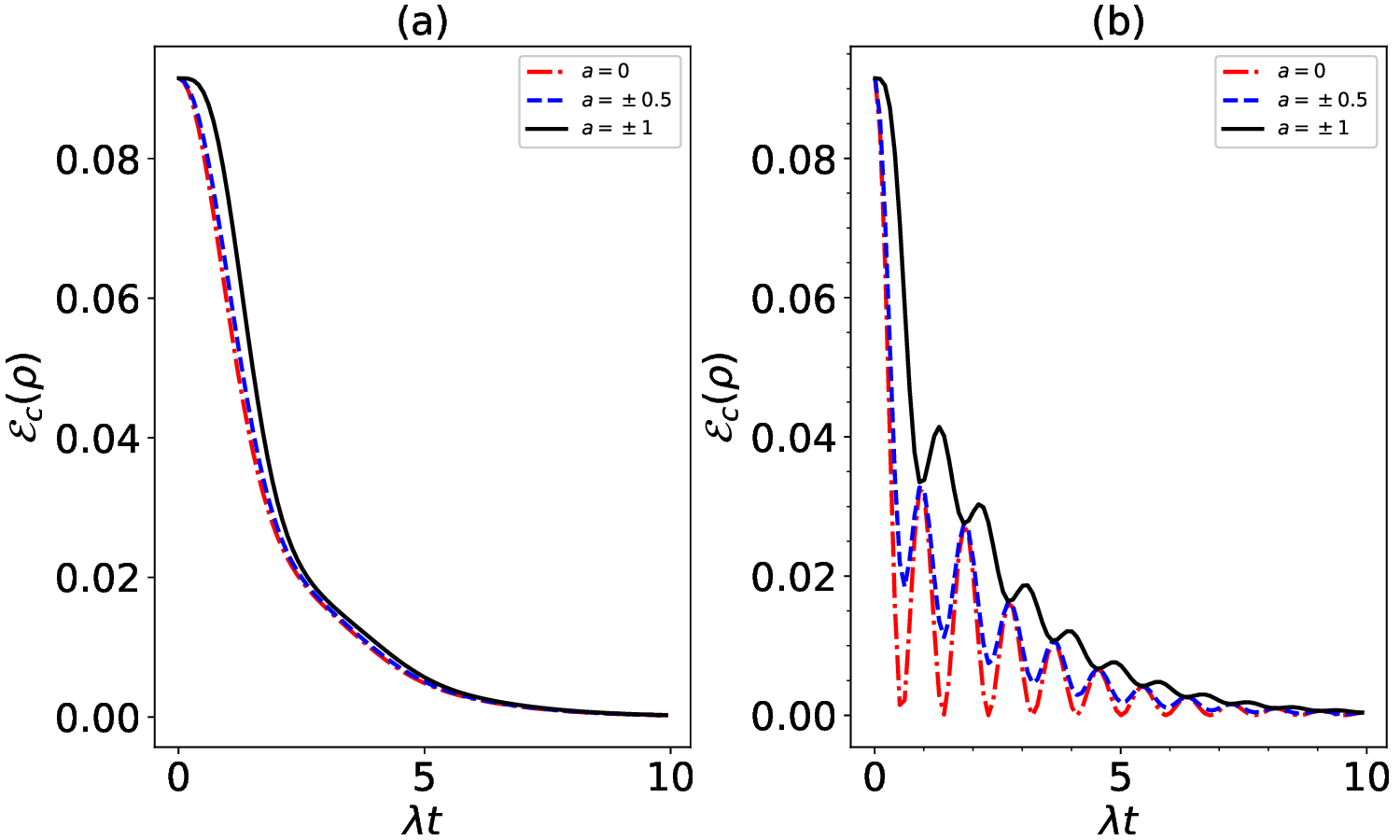}
    \centering
    \caption{(a)Coherent part of the ergotropy  $\mathcal{E}_c(\rho)$   as a function of dimensionless parameter $\lambda t$ for different amount of the non-equilibrium parameter  $a$ with $r_x=r_y=r_z=0.25$ in (a) Markovian regime, $v/\lambda=0.8$ and $\kappa/\lambda =2$. (b)non-Markovian regime, $v/\lambda=3$ and $\kappa/\lambda =2$.}\label{Fig3}
\end{figure}
Fig.\ref{Fig3}, shows the coherent part of ergotropy as a function of dimensionless parameter $\lambda t$ for different values of  non-equilibrium parameter $a$ with initial Bloch vector component $r_x=r_y=r_z=0.25$. In Fig.\ref{Fig3}(a), the dynamics of coherent part of ergotropy in Markovian regime has been shown and Fig.\ref{Fig3}(b) shows the dynamics of coherent part of ergotropy in non-Markovian regime. It can be observed that for both Markovian and non-Markovian regime the coherent part increases with increasing the non-equilibrium parameter $a$. t is also observed that the coherent part of the ergotropy asymptotically approaches zero. This is due to the nature of the noise, which affects only the coherence of the quantum system and leads to its complete destruction i.e. $\lim_{t \to \infty}\mathcal{E}_c(\rho)=0$. his final result is evident from Eq.\ref{cop}, where setting the coherence to zero, $C_{l_1}(\rho)=0$ leads to zero value for coherent part of the ergotropy $\mathcal{E}_c(\rho)$.

\section{SUMMARY AND RESULTS}
Ergotropy represents the maximum amount of work that can be extracted from a quantum system by employing suitable quantum operations. Unlike classical thermodynamics, where work extraction is based solely on energy differences between states, ergotropy considers the potential for work extraction from quantum coherence. In this work, we investigated the dynamics of ergotropy in the presence of a non-equilibrium environment. The considered non-equilibrium noise primarily affects the coherence of the quantum system and not the population of energy levels. Therefore, the internal energy, which is determined by the populations of the energy levels (diagonal elements of the density matrix), remains largely unaffected by noise. So, due to the effect of this non-equilibrium environment on the system, the coherent part of the ergotropy vanishes over time and just its incoherent part of ergotropy unaffected. Furthermore, it was observed that increasing the non-equilibrium parameter leads to an increase in the amount of the ergotropy. In other words, it can be concluded that the extraction of work from the quantum system in a non-equilibrium environment occurs more optimally compared to an equilibrium environment.


\end{document}